# Ferrovalley Physics in Stacked Bilayer Altermagnetic Systems

Yun-Qin Li,[1] Yu-Ke Zhang,[1] Xin-Le Lu,[1] Ya-Ping Shao,[1] Zhi-qiang Bao,[1] Jun-Ding Zheng,[1, *] Wen-Yi Tong,[1, †] Chun-Gang Duan[1,2, §]

[1]Key Laboratory of Polar Materials and Devices (MOE), School of Physics and Electronic Science and Shanghai Center of Brain-inspired Intelligent Materials and Devices, East China Normal University, Shanghai 200241, China
[2]Collaborative Innovation Center of Extreme Optics, Shanxi University, Taiyuan, Shanxi 030006, China

As an emerging magnetic phase, altermagnets with compensated magnetic order and non-relativistic spin-splitting have attracted widespread attention. Currently, strain engineering is considered to be an effective method for inducing valley polarization in altermagnets, however, achieving controllable switching of valley polarization is extremely challenging. Herein, combined with tight-binding model and first-principles calculations, we propose that interlayer sliding can be used to successfully induce and effectively manipulate the large valley polarization in altermagnets. Using $Fe_2MX_4$ (M = Mo, W; X = S, Se or Te) family as examples, we predict that sliding induced ferrovalley states in such systems can exhibit many unique properties, including the linearly optical dichroism that is independent of spin-orbit coupling, and the anomalous valley Hall effect. These findings imply the correlation among spin, valley, layer and optical degrees of freedom that makes altermagnets attractive in spintronics, valleytronics and even their crossing areas.

## I. INTRODUCTION

Altermagnets (AMs), an emerging magnetic phase, have compensated magnetic order in real space but exhibit the non-relativistic spin-splitting in the reciprocal space, which are receiving widespread attention and research fervor [1-10]. AMs with novel physical properties, such as the crystal Hall effect, spin currents and torques, and tunneling magnetoresistance, have great potential to lead technological innovations in spintronics fields for the fabrication of high-capacity storage devices [11-15]. Up to now, many AMs have been theoretically predicted and experimentally verified, such as bulk MnTe, $RuO_2$ and CrSb [16-22]. In addition to charge and spin, the valley degree of freedom in valleytronic materials serves as information carriers and are widely utilized for information encoding or storage [23-26]. Ferrovalley materials with spontaneous valley polarization has been firstly proposed in transition metal dichalcogenides in presence of spin-valley locking [27]. Up to now, many spin-related ferrovalley systems have been predicted, such as $FeCl_2$, VClBr and $VSiGeP_4$ [28-32], encouraging the fusion of valleytronics with spintronics. Very recently, the valley-polarized states have been proposed in AMs, including the $V_2Se_2O$, $Cr_2O_2$ and $A(BN)_2$ [33-42], which introduces new members to the ferrovalley family. However, valley polarization (VP) in these AMs is mainly realized by strain engineering [33-42], posing significant challenges for experimental control. In view of this, the discovery of other effective methods to regulate VP in AMs is urgent and challenging.

Previous studies proved that the sliding engineering opens a new chapter in the study of the introduction and manipulation of VP in two-dimensional materials [43-47]. Inspired by these, we predict that the interlayer sliding can also be a practical method to induce and control the valley polarization in AMs in addition to strain engineering. We demonstrate this mechanism via first-principles calculations and tight-binding model analysis in the concrete bilayer $Fe_2MX_4$ family. In such altermagnetic systems with VP induced by sliding effect, there exist intriguing properties, including the spin-valley locking and the anomalous valley Hall effect. Interestingly, the optical selection rule is found at valleys, which has not been discussed in previously reported AMs systems [33-42], offering a noncontact and nondestructive method to detect the emergence of VP and reversal of its polarity. Our findings realize the modulation of valley degree of freedom in AMs by interlayer sliding engineering, offering new perspectives and possibilities for exploring the potential applications of AMs, and paving the way for the development of advanced multifunctional nanodevices in the crossing fields of spintronics and valleytronics.

## II. RESULTS

The fully relaxed $Fe_2MX_4$ structure with the X-M-X sandwich structure is shown in Fig. 1(a), which possesses the space group of $P$-$42m$ (No. 111, $D_{2d}$ symmetry), same to that of $Cu_2MX_4$ and $V_2MX_4$ monolayers [48-50]. Due to the separation of two metal Fe ions by X ions, there is almost no overlap between the electron clouds of adjacent Fe ions, preventing direct exchange for occurring. In fact, the direct exchange of magnetic Fe ions with X ions leads to the indirect exchange interaction between Fe$_1$ and Fe$_2$ ions (in Fig. 1(b)). Such superexchange coupling effect between Fe$_1$ and Fe$_2$ makes them antiferromagnetic (AFM) aligned, in consistent to the computed AFM magnetic ground state with lowest energy (in Fig. S1). The spin density shows that the *PT* symmetry of magnetization density on the opposite Fe spin sublattices is broken by the crystal arrangement of W and X atoms. Two sublattices are linked by mirror symmetries with respect to Fe-X-Fe plane ($\sigma_d$ operations), rather than crystal translation or inversion (in Fig. S2). Therefore, Fe$_2$MX$_4$ is regarded as a type-I altermagnet [10], where its spin splitting is independent of spin-orbit coupling (SOC). As expected, band structures of Fe$_2$MX$_4$ exhibit the noticeable non-relativistic spin splitting (in Figs. 1(c) and S3). Calculations of phonon spectra (in Fig. S4) and elastic coefficients $C_{kj}$ (in Table SII) indicate the dynamic and mechanical stabilities of the Fe$_2$MX$_4$ systems, which is in consistent to previous study [51]. Indeed, the Cu$_2$MX$_4$ materials with the same structure have been experimentally synthesized [48,49], implying the high probability to fabricate Fe$_2$MX$_4$ films.

*jdzheng@phy.ecnu.edu.cn
†wytong@ee.ecnu.edu.cn
§cgduan@clpm.ecnu.edu.cn

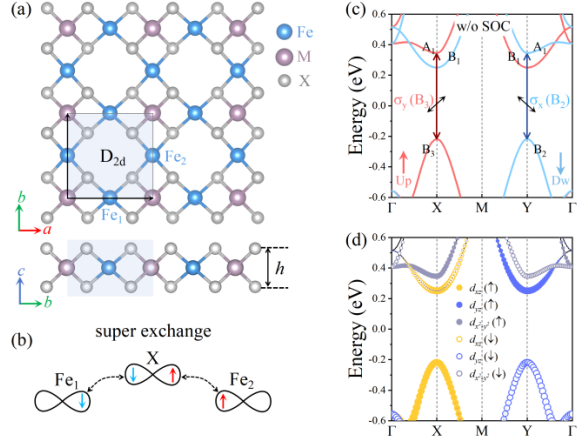

FIG. 1. (a) Top and side views of Fe$_2$MX$_4$ monolayers. (b) The superexchange interactions between Fe$_1$ and Fe$_2$ atoms via X atoms. (c) Spin-polarized and (d) *d*-orbital projected band structures of Fe atoms in Fe$_2$WTe$_4$ monolayer without consideration of SOC. Irreducible representations of bands, *x*- and *y*-polarized lights are marked in (c). The up and down arrows in (d) represent the spin-up and spin-down channels of Fe atoms, respectively.

As shown in Figs. 1(d) and S5, the valence band maximum (VBM) at X and Y valleys is primarily contributed by $d_{xz}^\uparrow$ and $d_{yz}^\downarrow$ orbitals of the Fe$_2$ and Fe$_1$ atoms in Fe$_2$WTe$_4$. While the conduction band minimum (CBM) at X and Y valleys is contributed by the $d_{xz}^\downarrow$ and $d_{yz}^\uparrow$ of the Fe$_1$ and Fe$_2$ atoms. Due to the prohibition of spin flipping, the optical transition between VBM and CBM at X and Y valleys is forbidden. However, we note that the second conduction band minimum (CBM-2), consisting of $d_{x^2-y^2}$ orbitals, possesses the same spin states as the VBM at a certain valley, making the optical transition between VBM and CBM-2 possible. There exists the spin-valley locking within Fe$_2$MX$_4$ system, where the X and Y valleys are locked with spin-up and spin-down channels respectively. According to the great orthogonality theorem, the irreducible representations (IRs) of VBM and CBM-2 meet B$_3$ ⊗ B$_3$ = A$_1$ in the X valley and B$_2$ ⊗ B$_2$ = A$_1$ in the Y valley (Details are provided in Supplemental Material (SM)). Such optical selection rule, i.e., the X and Y valleys can independently absorb the *y*- and *x*-polarized light, which hasn't been explored in previous reported altermagnets [33-42]. The imaginary parts of complex dielectric function ε$_2$ of Fe$_2$MX$_4$ calculated by our OPTICPACK package [27] are in line with this point. As shown in Fig. S6, absorption edge of *y*- and *x*-polarized light correspond to the energy gap between VBM and CBM-2, proving the prohibition of spin flipping from VBM-CBM hopping. The equivalent energy gap between VBM and CBM-2 in X and Y valleys, along with the degenerated absorption spectra excited by *y*- and *x*-polarized light, demonstrate the paravalley characteristics without spontaneous VP of the pristine Fe$_2$MX$_4$. In order to induce VP, strain engineering, which has been adopted in previous valley-related studies of altermagnets [33-41], might be a viable approach. Young's modulus and Poisson's ratio reveal the anisotropic and flexible mechanical properties of Fe$_2$MX$_4$ (see Fig. S7), benefiting to modulates physical properties by strains. As shown in Figs. S8 and S9, our calculations of electronic structures and optical spectra prove the effectiveness of inducing VP by uniaxial strain. Nevertheless, controlling VP is still challenging, in demand of switching strain direction between the *x*- and *y*-axis. More effective methods are thus expected to produce and modulate VP.

Sliding engineering is successful in introducing and manipulating VP of multilayers [43-47]. Similar strategy is then employed in Fe$_2$MX$_4$ bilayers. Firstly, we build a simplified effective tight-binding (TB) model for Fe$_2$MX$_4$ monolayer with $D_{2d}$ point group. We here only consider the $d_{xz}^\downarrow$, $d_{yz}^\downarrow$, $d_{x^2-y^2}^\downarrow$ and $d_{xz}^\uparrow$, $d_{yz}^\uparrow$, $d_{x^2-y^2}^\uparrow$ for Fe$_1$ and Fe$_2$ atoms. The schematic illustration of hopping between lattice sites with the consideration of the nearest-neighbor terms in our TB model is shown in Fig. S10. The hopping interactions between two Fe atoms can be ignored due to their opposite spin states [52], making the zero

values of anti-diagonal terms. The Hamiltonian $H(k)$ for $Fe_2MX_4$ monolayer can be written as:

$$H(k) = \begin{bmatrix} H^{\downarrow}(k) & 0 \\ 0 & H^{\uparrow}(k) \end{bmatrix} \quad (1)$$

$$H^{\downarrow(\uparrow)}(k) = \begin{bmatrix} H^{\downarrow(\uparrow)}_{d_{xz}} & 0 & 0 \\ 0 & H^{\downarrow(\uparrow)}_{d_{yz}} & 0 \\ 0 & 0 & H^{\downarrow(\uparrow)}_{d_{x^2-y^2}} \end{bmatrix} \quad (2)$$

$$H^m_i = \varepsilon^m_i + t^m_{1i}\cos(kx) + t^m_{2i}\cos(ky),$$
$$(m = \downarrow, \uparrow; i = d_{xz}, d_{yz}, d_{x^2-y^2}) \quad (3)$$

where $k$ represents the wave vector, $\varepsilon$ and $t$ are on-site energy and hopping parameter, respectively. As shown in Fig. S11, the band structures derived from the Hamiltonian qualitatively rebuild our DFT results for the intrinsic and strained $Fe_2MX_4$, which verifies the effectiveness of our six orbitals TB model. We then extend the TB model to $Fe_2MX_4$ bilayers (in Fig. 2(a)), where the interlayer nearest neighbor hopping is taken into account. The Hamiltonian $H'(k)$ can be written as:

$$H'(k) = \begin{bmatrix} H^{11}(k) & H^{12}(k) \\ H^{21}(k) & H^{22}(k) \end{bmatrix} \quad (4)$$

where the superscripts 1 and 2 represent the bottom and top layers. The $Fe_2WX_4$ bilayers are intralayer antiferromagnetic and interlayer ferromagnetic couplings, i.e. $H^{11}(k) = H^{22}(k) = H(k)$. Details of TB model is shown in SM. As illustrated in Fig. S12, for the $AC_1$ ($AC_2$) state, there is $H^{12}(k) = t'\cos(kx)$ ($H^{12}(k) = t'\cos(ky)$), indicating the breaking of $\sigma_d$ symmetry (The $t'$ is the hopping between same orbital with same spin in two monolayers). Since all the symmetry operations connecting two valleys are absent, VP state with inequivalent band gap at X and Y valleys is expected to be introduced [46]. In addition, the Hamiltonian of $AC_1$ and $AC_2$ states can be regarded as the index exchange between $x$ and $y$, the sign of VP for these two states should thus be reverted. Switching between $AC_1$ and $AC_2$ can be realized through the intermediate AB state. It preserves mirror symmetries along Fe-X-Fe planes, corresponding to the pristine paravalley state without spontaneous VP. Band structures of bilayer $Fe_2WX_4$ obtained from our TB model (Fig. S12) indicate that interlayer sliding is indeed an effective strategy to introduce and reverse VP through operating $\sigma_d$ symmetries.

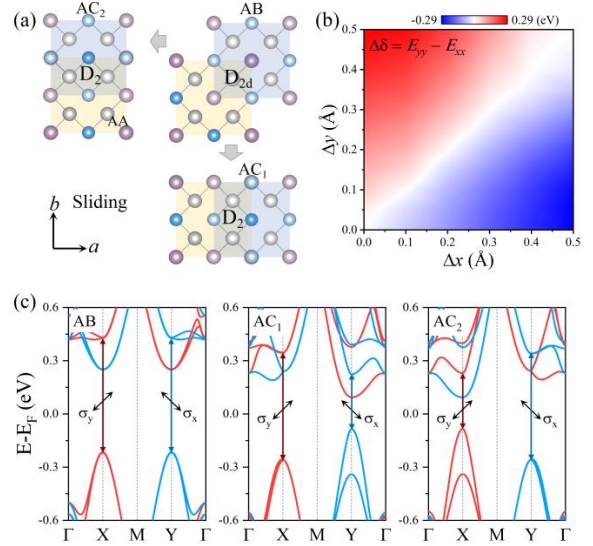

FIG. 2. (a) Schematic structure of a $Fe_2MX_4$ bilayer with AB, $AC_1$ and $AC_2$ stacking orders. (b) Valley polarization difference contour versus ($\Delta x$, $\Delta y$) of $Fe_2WTe_4$ bilayer. Red and blue region represent the optical absorption energy gap of Y valley ($E_{xx}$) is larger and lower than the X valley ($E_{yy}$), respectively. (c) Spin-projected band structures without SOC of $Fe_2WTe_4$ bilayer with AB, $AC_1$ and $AC_2$ states.

Following the TB models, we carry out DFT calculations for the $Fe_2WTe_4$ bilayers. When the top layer slides along the diagonal direction, $Fe_2WTe_4$ bilayers maintain the $D_{2d}$ symmetry, thus stabilizing in the pristine paravalley state with equivalent absorption edge of $y$- and $x$-polarized light (the upper panel of Fig. 3(a)) as the monolayer systems. When the sliding along the $x$- or $y$-axis direction occurs, band structures of $Fe_2WTe_4$ bilayer show inequivalent energy gaps at X and Y valleys (in Fig. 2(c)), corresponding to the ferrovalley state with VP of 0.29 eV (in Fig. 2(b)). Their optical properties own the linear dichroism (in Fig. 3(a)). In the ferrovalley $AC_1$ state, compared with the spectrum excited by $y$-polarized light at X valley, the $x$-polarized one at Y valley experiences a red shift (the middle panel of Fig. 3(a)). When it slides to $AC_2$ state, as clearly displayed in Fig. 2(c), its VP possess reversed polarity that energy gap in X valley is smaller than the one in Y valley. Following this, in comparison with the $y$-polarized one, the blue shift of absorption edge related to $x$-polarized light happens (the bottom panel of Fig. 3(a)). These results indicate that the VP state in $Fe_2WTe_4$ bilayers can be noncontactively and nondestructively detected by optical methods, promoting the practical utilization of valley degrees of freedom in AMs for information storage and coding.

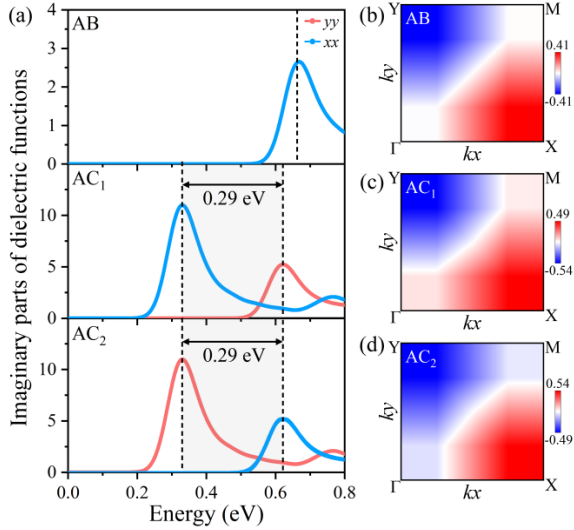

FIG. 3. (a) The imaginary parts of complex dielectric function $\varepsilon_2$ and (b-c) Contour maps of Berry curvatures in units of Å$^2$ in 2D Brillouin zone in Fe$_2$WTe$_4$ bilayers.

Previous analysis in the absence of SOC effect proves that sliding engineering is promising in inducing and manipulating VP of altermagnets. We then explore the effect of SOC on valley physics in Fe$_2$MX$_4$ systems. As shown in Figs. S13-S16, the effect of SOC on band dispersion for Fe$_2$MX$_4$ monolayer and bilayer systems is negligible, and the optical selection rule is strongly robust against SOC. The presence of SOC may provide interesting phenomena related to Berry curvatures $\Omega(k)$ [27]. We then calculate the $\Omega(k)$ for the VBM in Fe$_2$MX$_4$ monolayers and bilayers. For monolayer and AB-stacked bilayer Fe$_2$WTe$_4$ that are paravalley states protected by $\sigma_d$ symmetry, the $\Omega(k)$ around X and Y valleys have opposite sign but identical absolute value $\Omega_z(X) = -\Omega_z(Y)$ (in Figs. 3(b) and S17). The existence of the valley Hall effect, characterized by long-lived spin and valley accumulations, can be expected there. Due to the mirror symmetry breaking in strained monolayers, AC$_1$ and AC$_2$ bilayers, the $\Omega(k)$ near X and Y valleys still has opposite sign, but their absolute values are different, indicating the introduction of VP, corresponding to the ferrovalley states. The reversal of VP caused by sliding leads to the inversion of the absolute values of $\Omega(k)$ in X and Y valleys, but their signs remain the same. For the AC$_1$ state (in Fig. 3(c)), the absolute value of $\Omega(k)$ in Y valley is larger than that of X valley, resulting in a negative summation. When the VP is reversed by sliding to the AC$_2$ state (in Fig. 3(d)), the summarized $\Omega(k)$ remains the same magnitude yet with opposite sign. Instead of valley Hall effect, the anomalous valley Hall effect (AVHE) exists in these ferrovalley states.

For ferrovalley Fe$_2$MX$_4$ bilayers, hole doping shifts the Fermi energy level crossing the top of valence band at Y or X valleys (in Fig. 4(a)). In the p-type AC$_1$ state (in Fig. 4(b)), the majority carriers, that is, spin-up holes from the Y valley, gain transverse velocities of $v_\perp \propto -E \times \Omega_z(k)$ [53,54] towards right side in the presence of external electric field. The accumulation of holes in the right boundary of the ribbon generates a charge Hall current that can be detected as a positive voltage. Conversely, spin-down holes from X valley will accumulate in the left side of ribbons consisting of the AC$_2$ stacked Fe$_2$MX$_4$ bilayer with reversed VP, resulting in a negative transverse charge current. Obviously, Berry curvature serves as an alternative way for determining the emergence of VP and its polarity in AMs. AVHE correlated with it makes the electrically reading and mechanically writing memory devices possible. The binary information is stored by the valley polarization of the ferrovalley Fe$_2$MX$_4$ bilayers with the advantage of nonvolatility that could be controlled by sliding through mechanical force. In addition, it can be easily 'read out' utilizing the sign of the transverse Hall voltage.

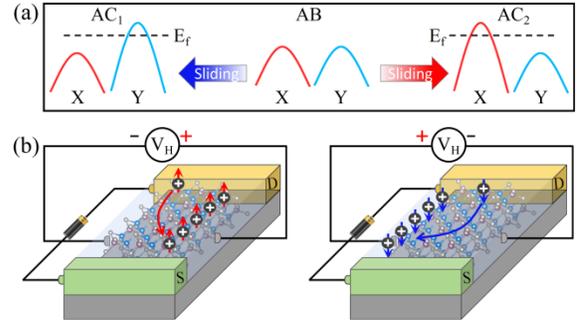

FIG. 4. (a) Sliding induced VP and the Fermi energy levels shift to the top of valence band at Y(X) valleys by holes doping. (b) Diagram of the anomalous valley Hall effect in stacked AC$_1$ and AC$_2$ bilayers under holes doping in presence of an in-plane electric field $E$.

### III. CONCLUSIONS

To summarize, based on the combination of the tight-binding model with first-principles calculations, we introduce that sliding engineering is a viable approach to not only induce but also manipulate VP in altermagnets. Taking the Fe$_2$MX$_4$ family as representatives, we demonstrate that the ferrovalley states generated by sliding will exhibit a variety of distinctive characteristics, including large VP, linearly optical dichroism and AVHE. We strongly advocate more theoretical and experimental efforts on such an ideal platform to study the interaction among layer, spin, valley, and optical degrees of freedom. It is of great importance in extending the research interest of altermagnets to valleytronic field and paving the way to their practical applications in the crossing area of spintronics and valleytronics.

**ACKNOWLEDGMENTS**

This work was supported by the National Key Research and Development Program of China (Grants No. 2022YFA1402902 and No. 2021YFA1200700), the National Natural Science Foundation of China (Grants No. 12134003 and No. 12304218), National funded postdoctoral researcher program of China (Grant No. GZC20230809), Shanghai Science and Technology Innovation Action Plan (Grant No. 21JC1402000), Shanghai Pujiang Program (Grant No. 23PJ1402200), and East China Normal University Multifunctional Platform for Innovation.